\documentclass[12pt]{iopart}

\usepackage{graphicx}

\usepackage{algpseudocode}
\usepackage[ocgcolorlinks,colorlinks=true,linkcolor=blue,citecolor=red]{hyperref}

\usepackage{iopams}

\usepackage{verbatim}

\newcommand{\eqref}[1]{(\ref{#1})}

\newcommand{\beq}{\begin{equation}}
\newcommand{\eeq}{\end{equation}}
\newcommand{\bea}{\begin{eqnarray}}
\newcommand{\eea}{\end{eqnarray}}

\newcommand\coloneqq{:=}
\newcommand{\pvec}[1]{\vec{#1}\mkern2mu\vphantom{#1}}

\def\erf{\mathrm{erf}}
\def\erfc{\mathrm{erfc}}

\begin{document}

\title[On the extreme value statistics of normal random matrices and 2D Coulomb gases]{On the extreme value statistics of normal random matrices and 2D Coulomb gases: Universality and finite $N$ corrections}

\author{R.~Ebrahimi$^1$ and S.~Zohren$^2$}

\address{
$^1$ Department of Physics, Pontifica Universidade Cat\'olica, Rio de Janeiro, Brazil \\
$^2$ Department of Materials, Advanced Research Computing, University of Oxford, UK
}



\begin{abstract}
In this paper we extend the orthogonal polynomials approach for extreme value calculations of Hermitian random matrices, developed by Nadal and Majumdar [1102.0738], to normal random matrices and 2D Coulomb gases in general. Firstly, we show that this approach provides an alternative derivation of results in the literature. More precisely, we show convergence of the rescaled eigenvalue with largest modulus of a normal Gaussian ensemble to a Gumbel distribution, as well as universality for an arbitrary radially symmetric potential. Secondly, it is shown that this approach can be generalised to obtain convergence of the eigenvalue with smallest modulus and its universality for ring distributions. Most interestingly, the here presented techniques are used to compute all slowly varying finite $N$ correction of the above distributions, which is important for practical applications, given the slow convergence. Another interesting aspect of this work is the fact that we can use standard techniques from Hermitian random matrices to obtain the extreme value statistics of non-Hermitian random matrices resembling the large $N$ expansion used in context of the double scaling limit of Hermitian matrix models in string theory.
\end{abstract}

\maketitle

\section{Introduction}

Complex non-Hermitian random matrices were first investigated by Ginibre \cite{Ginibre}, who analysed matrices whose elements are independent and identically distributed (real or  complex) Gaussian variables. He showed that the eigenvalue density $\mu_{N}(z)$ of such $N$ dimensional random matrices
in the limit $N\to\infty$ is given by the well-known circular law \footnote{The eigenvalue density is formally defined as $\int_A\mu_{N}(z)\,\mathrm{d}^2z=\frac{1}{N}\,\mathbb{E}_N(\#\{\mbox{eigenvalues in A}\})$ for any $A\subset\mathbb{C}$, where $\mathbb{E}_N(\cdot)$ indicates the expectation value with respect to the joint probability density function $\mathbb{P}_N$ over the $N$ eigenvalues.}
\beq
\mu(z) := \lim_{N\to\infty}\mu_{N}(z) = \frac{1}{\pi}\,\mathbb{I}_{\{|z|\le1\}}.
\eeq
In other words, the eigenvalues are distributed uniformly over the disk of radius 1 in the complex plane (see Figure \ref{fig:EigenvalueDensities} (Left)). 

In recent years there has been an increasing interest in non-Hermitian random matrices (see e.g. \cite{Akemann1, Akemann2, Bordenave, Rourke, Feinberg} for some examples as well as Chapter 18 of \cite{Gernot-book} for a review). Random matrices where all elements are chosen independently from some distribution are generally referred to as Wigner matrices. The Gaussian case discussed above is a special case, since it allows for a Coulomb gas formulation describing the joint eigenvalues distribution in terms of an attractive Gaussian potential and a logarithmic, Coulomb repulsion \cite{Ginibre}. In general, Wigner random matrices do not allow for such a formulation due to the lack of symmetries. One closely related ensemble is that of normal random matrices \cite{Chau} where one introduces an additional symmetry that the matrices are normal (they commute with their adjoint). This symmetry allows for a Coulomb gas formulation with arbitrary potential. Another important class of non-Hermitian random matrices are isotropic or bi-unitarily invariant random matrices (see for example \cite{Sommers,Mario}).

Normal random matrices have been studied intensively during the last years \cite{ Veneziani, Ameur, Elbau, Weigmann}. In this work we are interested in the extreme value statistics of normal random matrices with general radially symmetric potential. As the eigenvalues can spread over the complex plane (within the support), they are less correlated than in the unitary or orthogonal case. This is reflected by the extreme value statistics. In the above example, considering the distribution of the eigenvalue with the largest modulus, it is known \cite{Rider, Chafai} that the rescaled distribution is given by a Gumbel distribution. Recall that the Gumbel distribution is one of three families which describe the extreme value statistics of independent and identically distributed random variables (Weilbull, Gumbel, Fr{\'e}chet) \cite{Gumbel, Haan, Kotz}. While the eigenvalues of non-Hermitian ensembles are not independent, their correlation is much weaker than for Hermitian ensembles. Indeed for Hermitian ensembles the eigenvalues are strongly correlated and the distribution of the largest eigenvalue is given by the Tracy-Widom distribution \cite{Tracy1,Tracy2,Tracy3}.

\paragraph{Outline and what's new.}

After a brief introduction to normal matrices and their relation to 2D Coulomb gases in Section \ref{sec:NRMT}, we extend in Section \ref{sec:Extreme} the orthogonal polynomials approach to extreme value statistics of random matrices from the Hermitian case \cite{Nadal} to non-Hermitian random matrices.
In Section \ref{sec:Gaussian} we use this approach to show convergence of the probability distribution of the rescaled eigenvalue with largest modulus of normal random matrices to a Gumbel distribution providing a simplified, alternative derivation of result in \cite{Rider,Chafai} for Coulomb gases. In Section \ref{sec:finite N} we show how the here presented approach can be used to calculate finite $N$ corrections to the Gumbel distribution. In particular, we provide an analytical expression including all slowly varying correction terms. 
In Section \ref{sec:outer} and \ref{sec:inner} we show that our approach can also be used to show convergence and universality of the distribution of the eigenvalue with largest modulus at the outer edge as well as the eigenvalue with smallest modulus at the inner edge of a ring distribution. In Section \ref{sec:finiteNgen} we derive finite $N$ results, analogous to those derived in Section \ref{sec:finite N} for the Gaussian, for the extreme value statistics at the outer and inner edge of arbitrary radially symmetric potentials. We conclude in Section \ref{sec:Discussion} where we also comment on future directions of research. At various places in this article analytical results are complemented with numerical results. The methods to obtain those numerical results are explained in \ref{sec:appA}. 

Note that while the results in Section \ref{sec:Gaussian} and \ref{sec:outer} are an alternative derivation of previously proven theorems \cite{Rider} and \cite{Chafai}, the here presented derivation is novel and interesting for several reasons. Firstly, we can use the same orthogonal polynomial approach as used for Hermitian random matrices in the case of non-Hermitian random matrices. Those techniques are widely used in the field of random matrix theory, thus making the derivation easily accessible. Secondly, it is non-trivial that we can use the same derivational steps as used to derive the Tracy-Widom distribution in \cite{Nadal} to show convergence of the extreme value statistics of normal random matrices to the Gumbel distribution. Furthermore, the derivation resembles the double scaling limit of the free energy in Hermitian matrix models \cite{Marino}. 

The results of Section \ref{sec:finite N}, \ref{sec:inner} and \ref{sec:finiteNgen} are new. Particularly interesting and important for practical applications is the analytical expression derived in Section \ref{sec:finite N} and \ref{sec:finiteNgen} for the finite $N$ contributions to the extreme value statistics including the Gumbel distribution and all slowly varying correction terms. Regarding the results for the inner edge, Section \ref{sec:inner}, it is interesting that one can use the same general approach as in Sections \ref{sec:Gaussian} and \ref{sec:outer}. However, alternatively one could have also used the mapping of eigenvalues $z_i \to 1/z_i$ and the corresponding change in the potential, $V(z) \to V(1/z) +2(1+1/N)\log|z|$, to derive the results of Section \ref{sec:inner} from Section \ref{sec:outer} or the corresponding result in the literature \cite{Chafai}.

\section{Normal random matrices and Coulomb gases} \label{sec:NRMT}

Normal random matrices were introduced in \cite{Chau} and are defined through the measure

\beq
\mathbb{P}_N(M)\,\mathrm{d}M=\frac{1}{N!\,Z_N}\,e^{-N\,\mathrm{tr}\,V(M)}\,\mathrm{d}M, 
\eeq
where $M$ are complex $N\times N$ dimensional matrices satisfying the constraint $[M,M^{\dagger}]=0$. Here  $\mathrm{d}M = \prod_{ij} \mathrm{d}\Re M_{ij}\, \mathrm{d}\Im M_{ij} $ is the so-called Lebesgue measure and the normalisation 
\beq
Z_N = \frac{1}{N!}\, \int e^{-N\,\mathrm{tr}\,V(M)}\,\mathrm{d}M
\eeq
is also referred to as the partition function.

\begin{figure}
  \centering
  \includegraphics[width=0.45\linewidth]{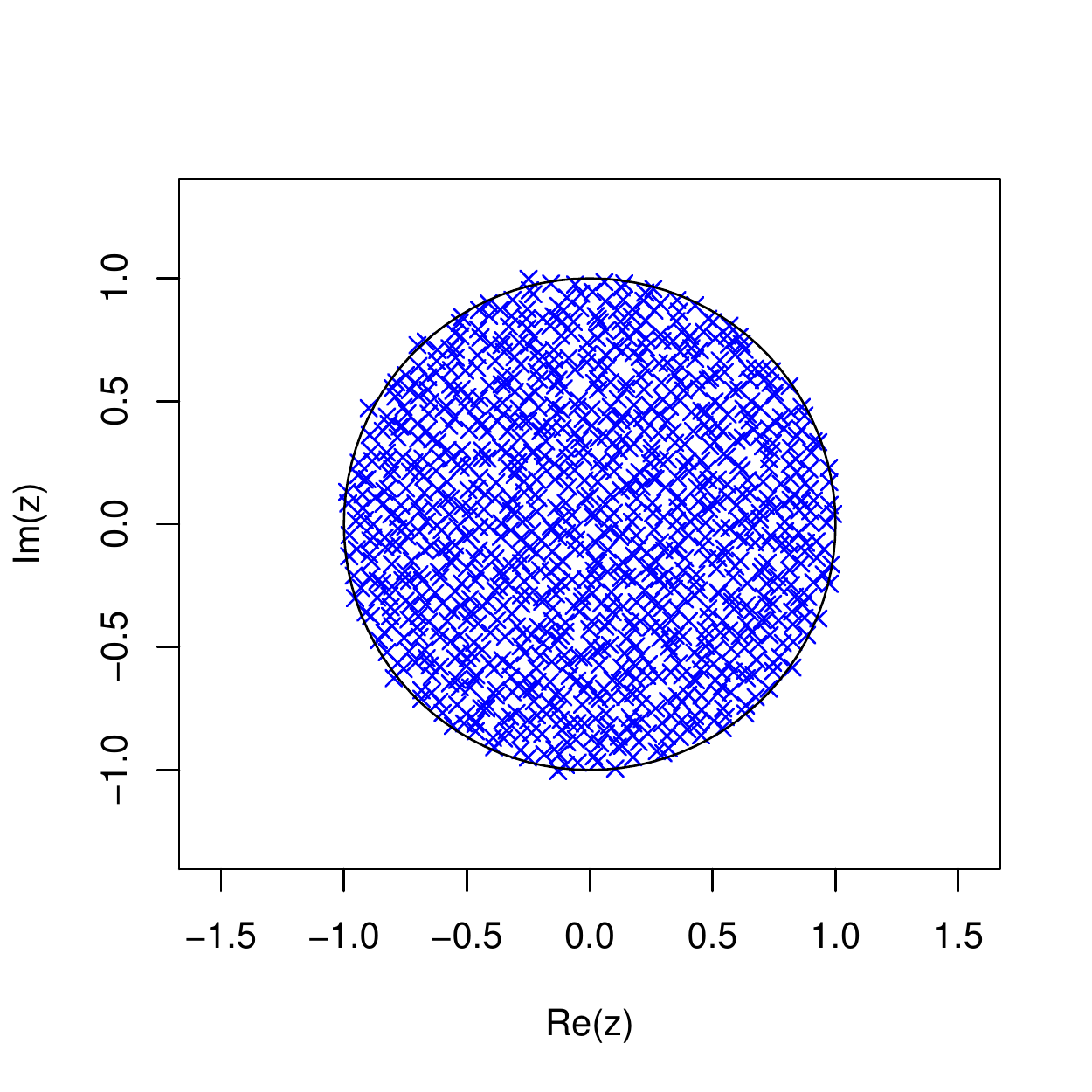}
   \includegraphics[width=0.45\linewidth]{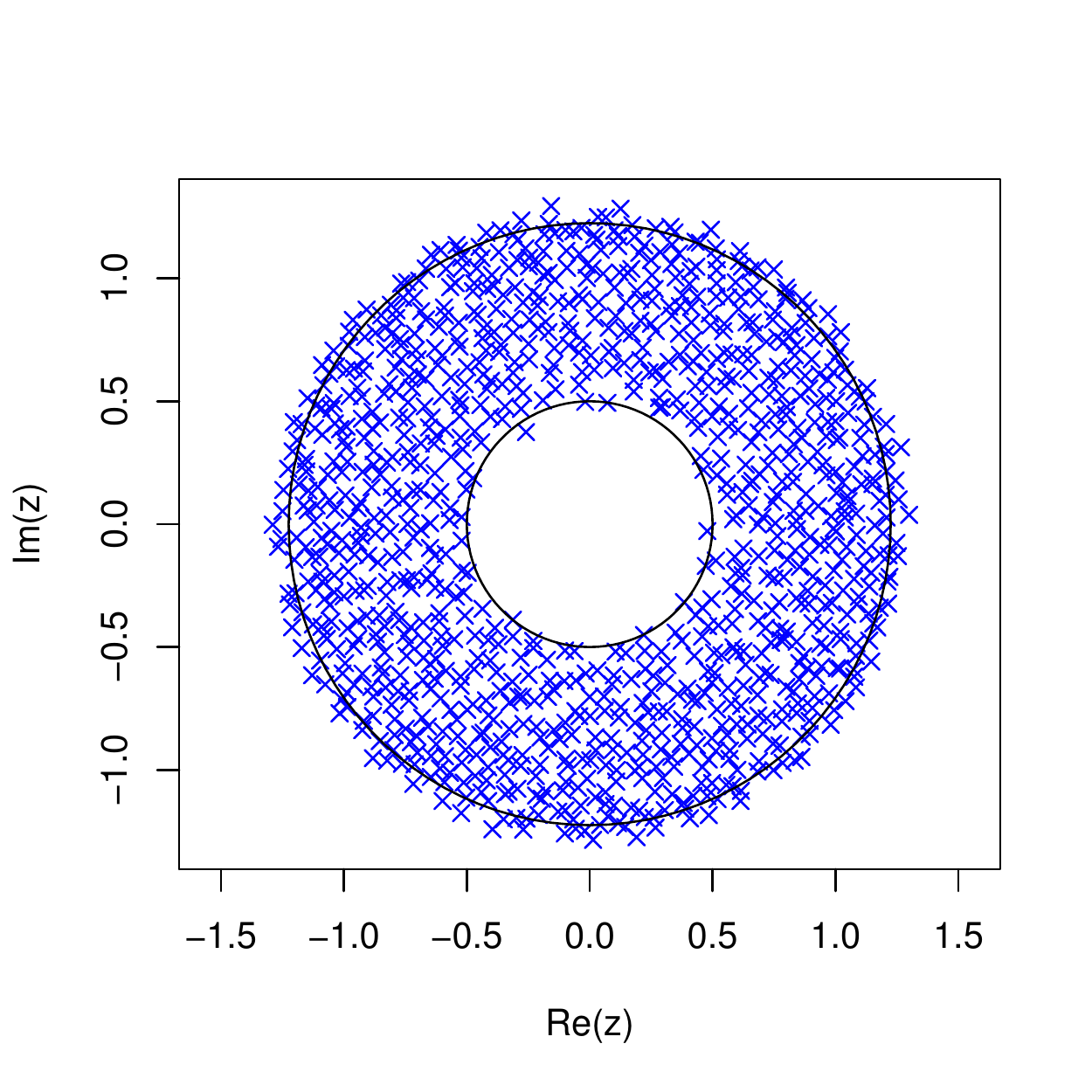}
  \caption{Left: Distribution of eigenvalues of a $1000\times1000$ Ginibre matrix equivalent to a Coulomb gas with Gaussian potential $V(r)=r^2$. The eigenvalues density is given by the circular law. Right: Distribution of eigenvalues of a $1000\times1000$ normal random matrix with potential $V(r)=r^2 - r$ obtained using Monte-Carlo methods. The eigenvalue density has support on a ring.}
  \label{fig:EigenvalueDensities}
\end{figure}

The eigenvalue statistics of normal random matrices is in many ways similar to that of general non-Hermitian matrices. However, normal random matrices are computationally much more feasible due to their symmetry. In particular, the normality condition $[M,M^{\dagger}]=0$ implies that there exists a unitary matrix which simultaneously diagonalises $M$ and $M^{\dagger}$. This allows one to employ Coulomb gas techniques and find the joint probability measure of the complex eigenvalues, given by
\beq\label{eq:Pmeasure}
\mathbb{P}_N(z_1,\ldots,z_N)=\frac{1}{N!\,Z_N}\,\prod_{1\le j<k\le N}|z_j-z_k|^2\,e^{-N\,\sum_{i=1}^NV(z_i)}.
\eeq
Here the term $\Delta(\{z\})\coloneqq\prod_{1\le j<k\le N}(z_j-z_k)$ is called the Vandermonde determinant. Writing
\beq \label{eq:Boltzmann}
\mathbb{P}_N(z_1,\ldots,z_N)=\frac{e^{-N^2 S_{\mathrm{eff}} (z_1,\ldots,z_N) }}{N!\,Z_N} 
\eeq
with
\beq
S_{\mathrm{eff}}(z_1,\ldots,z_N)=  \frac{1}{N} \sum_{i=1}^NV(z_i) -  \frac{2}{N^2} \sum_{1\le j<k\le N} \log | z_j-z_k |
\eeq
we can interpret the effective model as that of a gas of $N$ particles in a potential $V(z)$ experiencing a logarithmic repulsion. This is equivalent to a Coulomb repulsion in 2D, thus the name Coulomb gas.

As mentioned above, the (complex or real) Ginibre ensemble where all entries are (complex or real) Gaussian random variables also has a joint probability density function of its eigenvalues given by \eqref{eq:Pmeasure} with Gaussian potential. It is interesting to study normal random matrices as their joint eigenvalue distribution is described by a Coulomb gas with general potential $V(z)$.

In this article we are only interested in radially symmetric potentials
\beq
V(z)\equiv V(r),\hspace{.5cm}z=r\,e^{i\phi}.
\eeq
In this case one can use saddle point techniques or more mathematical results from the theory of harmonic analysis \cite{Saff} to find the eigenvalue density for a given radially symmetric potential. More precisely, given the condition that
 \beq\label{eq:bounded}
 \lim_{r\to\infty}\left(V(r)-\log r^2\right)=\infty
 \eeq
and $r\,V'(r)$ increasing in $\mathbb{R}^{+}$ or $V'(r)>0$ and $V$ convex in $\mathbb{R}^{+}$ one finds that the eigenvalue density is given by \cite{Saff}
\beq
\mu(z)\,\mathrm{d}^2z=\frac{1}{4\pi}\,\frac{\mathrm{d}}{\mathrm{d}r}\Big(r\,V'(r)\Big)\,\mathbb{I}_{\{a_-\le r\le a_+\}}\,\mathrm{d}r\,\mathrm{d}\phi.
\eeq

We see that the eigenvalue density has support on a ring with inner radius $a_-$ and outer radius $a_+$ (or a disk in the case where $a_-=0$), where the radii are given by
\beq\label{eq:inner}
V'(a_-)=0
\eeq
and $a_+$ is the solution to following equation
\beq\label{eq:outer}
a_+\,V'(a_+)=2.
\eeq

We see that the eigenvalue density either has topology of a disk or of an annulus. This is the famous one ring theorem \cite{Guionnet, Girko, Saff, Feinberg}. 

As an example, for the Gaussian normal ensemble with potential $V(r)=r^2$, we have $a_-=0$ and $a_+=1$, and the eigenvalues spread over the unit disk in the complex plane. We see that the eigenvalue density is identical to that of the Ginibre ensemble, i.e., the circular law (see Figure \ref{fig:EigenvalueDensities} (Left)). Another easy but more interesting example is the potential $V(r)=r^2+sr$. Here for $s>0$ the topology of the support is a disk while for $s<0$ the topology is that of an annulus (see Figure \ref{fig:EigenvalueDensities} (Right)). Interesting statistics can be computed for the transition when the topology changes from disk to annulus (see for example \cite{Cunden}).

Besides saddle point analysis, another powerful technique employed in random matrix theory in general and in this article in particular is the method of orthogonal polynomials. The aim of the orthogonal polynomials approach is to define a set of polynomials which are orthogonal with respect to a weight $w(z)$, i.e.,
\beq \label{eq:OPdef}
\langle p_n| p_m\rangle_w\coloneqq\int w(z)\,p_n(z)\,\overline{p_m(z)}\,\mathrm{d}^2z=h_n\,\delta_{nm}
\eeq
where in our case $w(z)=e^{-N\,V(z)}$. Having found such a set of polynomials one can use them to evaluate many quantities in random matrix theory, in particular, the partition function can be obtained as 
\beq
Z_N = \prod_{n=0}^{N-1} h_n,
\eeq
where $h_n$ are the normalisation constants of the orthogonal polynomials as defined in \eqref{eq:OPdef}.

For the case of radially symmetric potentials, the orthogonal polynomials become very simple. In fact, for any weight $w(z) = w(|z|)$ it can be readily checked that they are simply given by the monomials
\beq
p_n(z)= z^{n}.
\eeq
Thus one has 
\beq \label{eq:OPrelationw}
h_n = 2\pi  \int_0^\infty r^{2n + 1} w(r)\,\mathrm{d}r =  2\pi  \int_0^\infty r^{2n + 1} e^{-N V(r)}\,\mathrm{d}r.
\eeq
For the case of the Gaussian potential with $V(r)= r^2$ the normalisation constants are given by $h_n = \pi  N^{-n-1} \Gamma(n+1) $.

\section{Orthogonal polynomial approach for extreme value statistics of random matrices} \label{sec:Extreme}

One of the most well-known examples of extreme value statistics of random matrices is the Tracy-Widom distribution \cite{Tracy1,Tracy2,Tracy3} originally obtained for the GUE. 
The orthogonal polynomial approach to extreme value statistics of random matrices was first introduced in \cite{Nadal} to provide an easy derivation of the Tracy-Widom distribution for the GUE. It has then been extended to arbitrary potentials and multi-critical points \cite{Akemann} as well as large deviations \cite{Atkin}. Here we show how one can use the same formulation for normal random matrices with complex eigenvalues.  More precisely we are interested in the statistics of the complex eigenvalue $z_{max}$ with the largest modulus. The basic idea underlying this approach is simple. Let us define a partition function $Z_N(\mathcal{D})$ where the eigenvalues are restricted to a domain $\mathcal{D} \subseteq \mathbb{C}$,
\beq \label{eq:genZ}
Z_N(\mathcal{D}) = \frac{1}{N!}\,\int \prod_{1\le j<k\le N}|z_j-z_k|^2\,e^{-N\,\sum_{i=1}^NV(z_i)}\,\prod_j  \mathbb{I}_{z_j\in\mathcal{D}}\,\mathrm{d}^2 z_j ,
\eeq
where $\mathbb{I}_{\,C}$ is the indicator function which is 1 if $C$ is true and 0 otherwise. Then $Z_N(\mathbb{C}) \equiv Z_N$, i.e., when integrating over the entire domain we recover the conventional partition function. Given the above, it follows that
\beq
\mathbb{P}_N( \mbox{all eigenvalues in $\mathcal{D}$})=\frac{Z_N(\mathcal{D})}{Z_N}.
\eeq

We can use this relation to express extreme value statistics in terms of $Z_N(\mathcal{D})$. More precisely, let us choose $\mathcal{D} = \{ z : \,  |z| \leq y\}$ and denote as shorthand $Z_N(y):=Z_N(\{ z : \, |z|\leq y\})$ (and thus $Z_N(\infty) = Z_N$), then the cumulative probability function of the eigenvalue with the largest modulus is given by
\beq
F_N(y):=\mathbb{P}_N(|z_{max}|\le y)=\mathbb{P}_N( |z_1|\le y,\ldots,|z_N|\le y)=\frac{Z_N(y)}{Z_N(\infty)}.
\eeq

Following the analogous approach for the GUE \cite{Nadal}, the key idea is now to define orthogonal polynomials for the partition function $Z_N(y)$. In particular, we can find the orthogonal polynomials such that
\beq\label{eq:orth}
\int e^{-N\,V(z)}\,\mathbb{I}_{\{|z|\le y\}}\,p_n(z;y)\,\overline{p_m(z;y)}\,\mathrm{d}^2z=h_n(y)\,\delta_{nm}.
\eeq
We have
\beq
Z_N(y)=\prod_{n=0}^{N-1}h_n(y)
\eeq
and thus
\beq\label{eq:probability}
F_N(y)=\mathbb{P}_N(|z_{max}|\le y)=\prod_{n=0}^{N-1}\frac{h_n(y)}{h_n(\infty)}.
\eeq

As above, we are interested in the case of radially symmetric potentials $V(z)=V(|z|)$. In this case, we observe that \eqref{eq:orth} is equivalent to \eqref{eq:OPdef} with a weight $w(z)\equiv w(|z|; y)=e^{- N\,V(|z|)}\,\mathbb{I}_{\{|z|\leq y\}}$. Since $w(z)$ is radially symmetric, we know from the previous section that the orthogonal polynomials are simply given by the monomials $p_n(z;y)\coloneqq z^{n}$. We see that, while in general $p_n(z;y)$ is a function of $y$, here all $y$-dependence is in the normalization $h_n(y)$. Therefore, using equation \eqref{eq:OPrelationw} we obtain
\beq\label{eq:h}
h_n(y)=2\pi\,\int_0^y r^{2n+1}\,e^{-N V(r)}\,\mathrm{d}r,
\eeq
where the indicator function in the weight reduces the integration range of $r$ to $[0,y]$. 

Once $N$ becomes large the probability density $f_N(y) = F_N'(y)$ will be more and more centered around the edge of the support, $a_+$. More precisely, one expects $(|z_{max}| - a_{+}) \sim N^{-1/2}$.\footnote{Note that $N^{-1/2}$ is the natural distance scaling for $N$ eigenvalues in a finite 2D domain. However, for the Unitary ensembles it is known that the scaling changes at the edge. This is not expected here as the correlation is weaker (which is the reason why one obtains the Gumbel distribution).  Mathematically, one can see that this scaling is indeed correct from the terms $\sqrt{N}\,(y-r_0)$ in later equations, cf.~\eqref{eq:hfromerf}. It is however interesting to consider so-called multi-critical models where more derivatives in \eqref{eq:Taylorf} become zero, potentially leading to different scalings and universality classes, cf.~\cite{Akemann} for corresponding results for Unitary ensembles.}
Thus to obtain convergence to a limiting distribution one has to rescale $y$ to a ``continuum'' variable $Y$,
\beq \label{eq:scaling}
y= a_{+} + C N^{-1/2}\,( \alpha_N +\beta_N  Y),
\eeq
where $C$ is a constant and $\alpha_N,\beta_N$ are slowly varying functions.\footnote{Recall that a function $f(N)$ is called slowly varying (at infinity) if $\lim_{N\to\infty} f(aN)/f(N) = 1$ for all $a>0$. A typical example is $\log N$. Loosely speaking a slowly varying function grows slower than any power of $N$.} 
The constant $C$ is chosen merely for convenience and could equally be absorbed in $\alpha_N$ and $\beta_N$. Since $y\geq 0$, we have $ Y \geq-\alpha_N /\beta_N $. For the limiting distribution to have support on $Y\in(-\infty,+\infty)$, we thus require $\alpha_N /\beta_N \to \infty$ as $N\to \infty$.
This is enough to obtain the extreme value statistics related to the eigenvalue with the largest modulus for the Gaussian potential, as we will show in the next section.

\section{The Gaussian potential} \label{sec:Gaussian}

From the results of the previous section, more precisely \eqref{eq:probability} and \eqref{eq:h}, it follows that the cumulative probability function of the eigenvalue with largest modulus for the Gaussian random normal matrix is given by

\bea\label{eq:cpf}
F_N(y) &=& \mathbb{P}_N(|z_{max}|\le y) = \prod_{n=0}^{N-1}\frac{\int_0^y r^{2n+1}\,e^{-Nr^2}\,\mathrm{d}r}{\int_0^{\infty} r^{2n+1}\,e^{-Nr^2}\,\mathrm{d}r} \nonumber \\ 
&=& \prod_{n=0}^{N-1}\frac{\gamma(n+1,Ny^2 )}{\Gamma(n+1)}=\prod_{n=0}^{N-1}P(n+1,Ny^2) \label{eq:gammaratio}
\eea
in which $P(a,z):= \gamma(a,z)/\Gamma(a)$ is the regularised incomplete gamma function. Note that this result can also be obtained from \cite{Kostlan}, where it was shown that the moduli of eigenvalues for a Gaussian potential follow the distribution of independent and identically distributed $\chi^2$ random variables.

One can try to reduce the large $N$ behaviour of \eqref{eq:gammaratio} using the asymptotic expansion of the regularised incomplete gamma function, where both arguments become large, however, it is more instructive to calculate the asymptotic behaviour from first principles using a saddle point evaluation. In particular, such a derivation will generalise to arbitrary potentials, where we cannot find a closed form expression such as \eqref{eq:gammaratio}.

In the following we determine the large $N$ expansion of $h_n(y)$ using a saddle point expansion. Note that we can also obtain it from the so-called ``string equation'', a set of equations describing a recursion relation for the coefficients of the orthogonal polynomials, which is widely used in Hermitian matrix models \cite{Marino}.

To perform the asymptotic expansion using a saddle point evaluation we start by writing

\beq \label{eq:hwithf}
h_n(y)=2\pi\,\int_0^y e^{Nf(r)}\,\mathrm{d}r, \quad \mbox{with}\quad f(r)=\frac{2n+1}{N}\;\log r-r^2.
\eeq
We can now expand $f(r)$ around its maximum at $r_0=\sqrt{(2n+1)/(2N)}$ as
\beq\label{eq:Taylorf}
f(r)=f(r_0)-\frac{1}{2}\big|f''(r_0)\big|\,(r-r_0)^2+\mathcal{O}\left((r-r_0)^3\right),
\eeq
The integration in \eqref{eq:h} then reduces to Gaussian integrals over finite intervals, yielding error functions:
\beq \label{eq:hfromerf}
\frac{h_n(y)}{h_n(\infty)}=\frac{\erf(\sqrt{Nb}\,r_0)+\erf(\sqrt{Nb}\,(y-r_0))}{1+\erf(\sqrt{Nb}\,r_0)} + ...,
\eeq
where $b\equiv\big|f''(r_0)/2\big|=2$ for the Gaussian potential. Here ``$+...$'' refers to the corrections coming from the $\mathcal{O}\left((r-r_0)^3\right)$ term in \eqref{eq:Taylorf}. That those corrections are indeed of lower order can be verified later. 

Now using the Euler-Maclaurin summation formula, we look at the first two leading order terms in \eqref{eq:probability} when both $N$ and $n$ become large, while their ratio is given by $\frac{n}{N}=:\xi\in[0,1]$. One obtains
\beq\label{eq:EM}
\mathrm{log}\big(F_N(y)\big)\!=\!N\!\int_0^1g(\xi;y)\,\mathrm{d}\xi+\frac{1}{2}\big[g(1;y)-g(0;y)\big]\!+\!\mathcal{O}\!\left(N^{-1}\right)\!,
\eeq
where we defined
\bea
g(\xi;y) &:=&\log\frac{h_{[\xi N]}(y)}{h_{[\xi N]}(\infty)} \nonumber \\
&=& \log\left(\frac{\erf\Big(\sqrt{Nb}\,r_0(\xi)\Big)+\erf\Big(\sqrt{Nb}\,\big(y-r_0(\xi)\big)\Big)}{1+\erf\Big(\sqrt{Nb}\,r_0(\xi)\Big)}\right) + ... \label{eq:g}
\eea
in which $r_0(\xi)=\sqrt{\xi}+\mathcal{O}\left(N^{-1}\right)$. Recall the asymptotic expansion of the error function
\beq\label{eq:erf}
\erf(z)\equiv\frac{2}{\sqrt{\pi}}\,\int_0^z e^{-x^2}\,\mathrm{d}x=1-\frac{e^{-z^2}}{\sqrt{\pi}z}+\mathcal{O}\left(z^{-3}\,e^{-z^2}\right).
\eeq

Note that the argument of the first error function in the denominator of \eqref{eq:g} is not necessarily large if $\xi$ is close to zero. However, the asymptotics of \eqref{eq:g} is dominated by the second term
\beq \label{eq:g1}
g(\xi;y)\sim-\frac{1}{\sqrt{\pi}}\frac{e^{-Nb\,(y-r_0(\xi))^2}}{\sqrt{Nb}\,(y-r_0(\xi)) (1+\erf\big(\sqrt{Nb}\,r_0(\xi)) )},\hspace{.5cm}N\to\infty.
\eeq
Note that the scaling \eqref{eq:scaling} ensures that the argument of the second error function $\sqrt{Nb}\,\big(y-r_0(\xi)\big) \to \infty$ as $N\to\infty$ as long as the slowly varying function $\alpha_N\to\infty$. Furthermore, since $r_0(\xi)\to a_+$ as $\xi \to 1$, the dominant contribution in \eqref{eq:g1} comes from $\xi$ close to one, in which case we have
\beq \label{eq:gintermediate}
g(\xi;y)\sim-\frac{1}{2\sqrt{\pi}}\frac{e^{-Nb\,(y-r_0(\xi))^2}}{\sqrt{Nb}\,\big(y-r_0(\xi)\big)},\hspace{.5cm}N\to\infty.
\eeq
Therefore, the leading term on the right hand sight of \eqref{eq:EM} is given by
\beq\label{eq:EM'}
N\,\int_0^1g(\xi;y)\,\mathrm{d}\xi\sim-\frac{N}{2\sqrt{\pi}}\int_0^1\frac{e^{-Nb\,(y-r_0(\xi))^2}}{\sqrt{Nb}\,\big(y-r_0(\xi)\big)}\,\mathrm{d}\xi,\hspace{.5cm}N\to\infty.
\eeq

\begin{figure}
  \centering
  \includegraphics[width=0.70\linewidth]{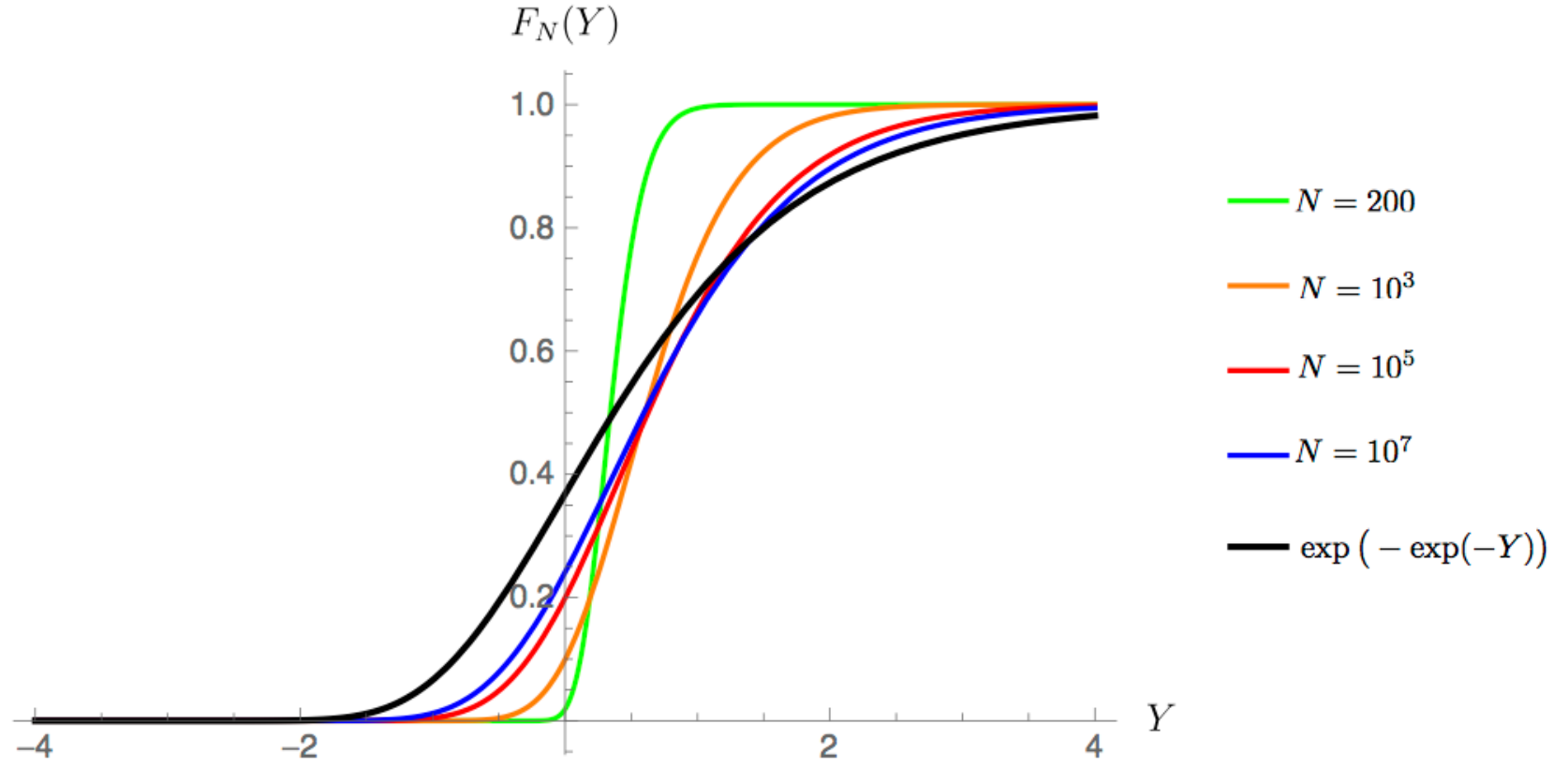}
  \caption{Shown is the function $F_N(Y)$ for increasing values of $N$, as well as the limiting case $F(Y) = \exp(-\exp(-Y))$. The function $F_N(Y)$ is obtained by numerically evaluating \eqref{eq:gammaratio} with $y$ replaced with its scaling relation \eqref{eq:yscalingrelation}.}
  \label{fig:finiteN}
\end{figure}

Let us now insert the scaling relation \eqref{eq:scaling}. We know that $\alpha_N$ is a slowly varying function which diverges for $N\to\infty$. Furthermore, we require $\alpha_N /\beta_N \to \infty$ as $N\to \infty$. It will become clear later that we will have to choose $\beta_N \sim 1/\alpha_N$ and moreover that it will be convenient to choose $\beta_N=1/(2\alpha_N)$. The factor of $2$ is merely a rescaling of $Y$ which could otherwise be determined at the end of the calculation. We thus have the scaling
\beq \label{eq:yscalingrelation}
y= a_+ + (Nb)^{-1/2} \left( \alpha_N + \frac{1}{2\alpha_N}Y \right),
\eeq
where here $a_+ =1$. Since we know that the contribution of $g(\xi;y)$ comes from $\xi$ close to one, we introduce the following change of variable in the integration in \eqref{eq:EM'},
\beq\label{eq:changev}
\xi=1-\frac{(2N)^{-1/2}}{\alpha_N } X.
\eeq
For this choice one has, recalling that $b=2$,
\beq\label{eq:errorarg}
\sqrt{Nb}\,\big(y-r_0(\xi)\big)=\alpha_N + \frac{1}{2\alpha_N }\,(X+Y)+\mathcal{O}\left(N^{-1/2}\alpha_N^{-2}\right).
\eeq
This relation also justifies the above choices of scalings. In particular, 

\beq\label{eq:square}
Nb\,\big(y-r_0(\xi)\big)^2 = \alpha_N^2 + (X+Y)+\mathcal{O}\left(\alpha_N^{-2}\right)
\eeq 
is a slowly varying function plus a finite term. To get this it was necessary to choose $\beta_N =1/(2\alpha_N)$. Putting everything together, one then obtains

\begin{eqnarray}\label{eq:integral}
N\,\int_0^1 g(\xi;y)\,\mathrm{d}\xi & \sim & -\frac{\sqrt{N}}{2\sqrt{2\pi}\,\alpha_N^2}\int_0^{\sqrt{2N}\,\alpha_N} e^{-\left[\alpha_N^2+(X+Y)\right]}\,\mathrm{d}X\nonumber\\
& \sim & -\frac{\sqrt{N}\,e^{-\alpha_N^2}}{2\sqrt{2\pi}\,\alpha_N^2}\,e^{-Y}.
\end{eqnarray}
For the pre-factor right-hand-side to converge to be 1, one is led to the following choice for the slowly varying function $\alpha_N$, to leading order,
\beq\label{eq:slowly}
\alpha_N=\frac{1}{\sqrt{2}}\,\big|\log N-2\log\log N-\log2\pi\big|^{\frac{1}{2}}.
\eeq
Using this, one arrives at

\beq\label{eq:largeN}
N\,\int_0^1 g(\xi;y)\,\mathrm{d}\xi=-e^{-Y}+\mathcal{O}\left(\log\log N/\log N\right).
\eeq
Furthermore, under this choice, for the subleading term on the right hand side of \eqref{eq:EM} we have

\beq\label{eq:sub}
\frac{1}{2}\,[g(1;y)-g(0;y)]=\mathcal{O} \left(e^{-\alpha_N^2} \alpha_N^{-1}\right)=\mathcal{O}\left(1/\sqrt{N\log N}\right).
\eeq
In addition, one can see that the corrections terms to \eqref{eq:hfromerf} are of the same order.

The final result is thus given by
\bea
F(Y)&\coloneqq&\lim_{N\rightarrow\infty} F_N(Y) \nonumber \\
&=&  \lim_{N\rightarrow\infty}\mathbb{P}_N\left(|z_{max}|\le1+(Nb)^{-1/2}\left[\alpha_N+\frac{Y}{2\alpha_N} \right]\right)=e^{\,-e^{-Y}}, \label{eq:finalGaussian}
\eea
which is precisely the Gumbel distribution. This provides an alternative derivation of the results in \cite{Rider}.

\begin{figure}
  \centering
  \includegraphics[width=0.45\linewidth]{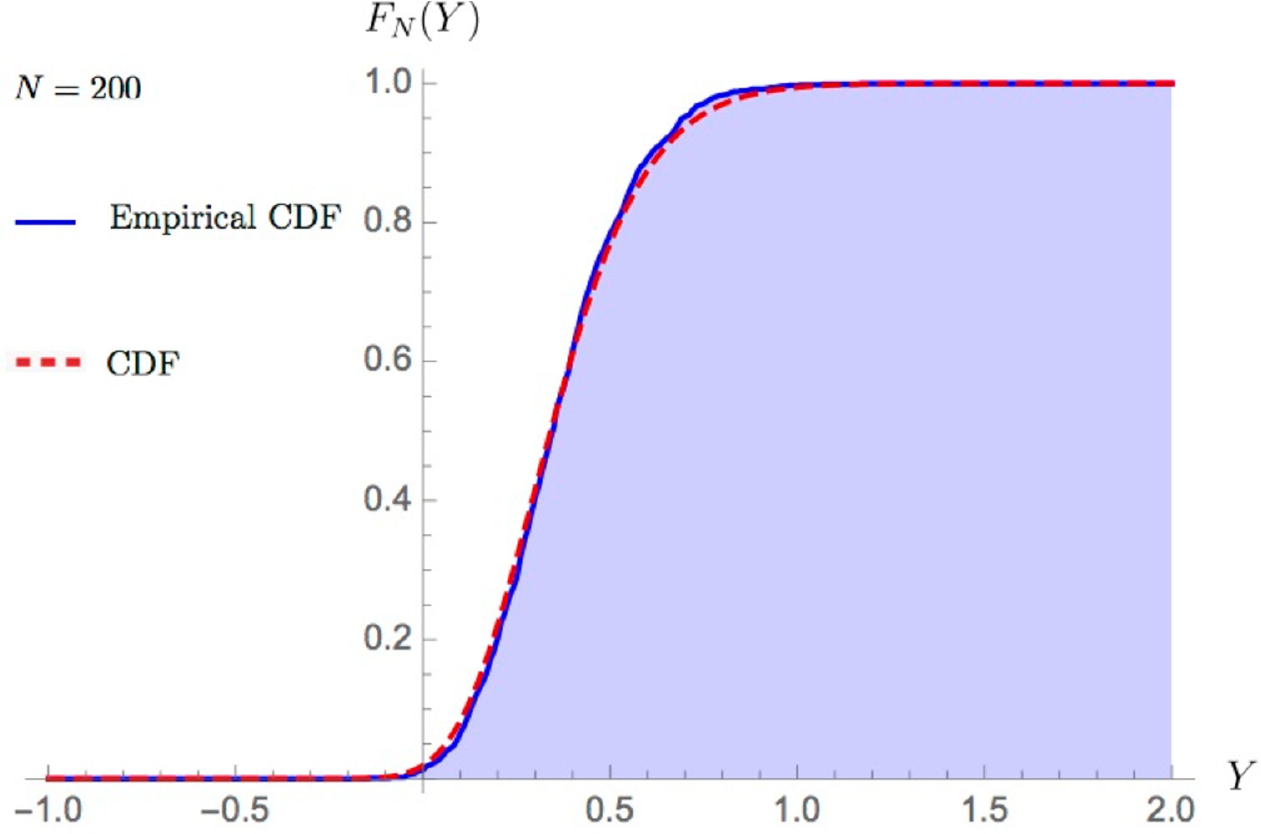}
   \includegraphics[width=0.45\linewidth]{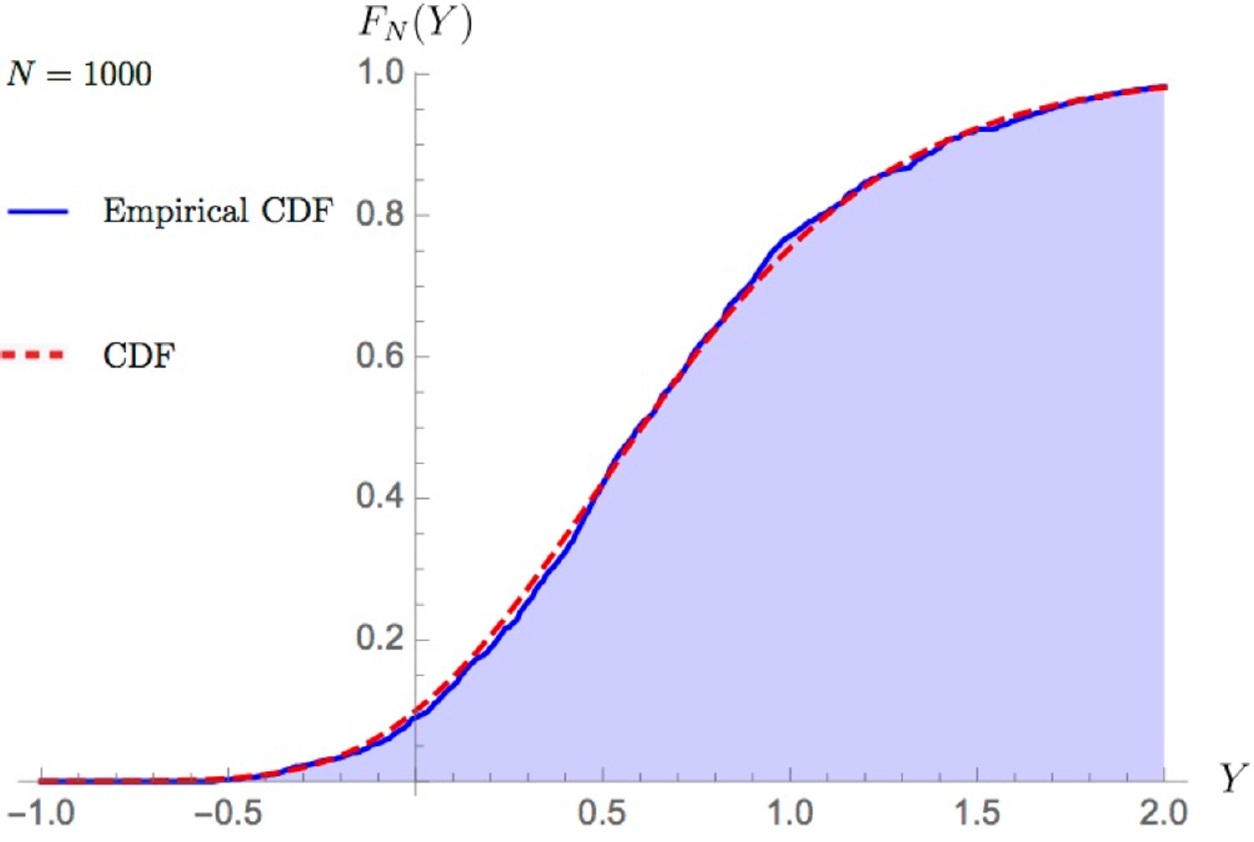}
  \caption{Comparison of the finite $N$ expression $F_N(Y)$ as given in \eqref{eq:gammaratio} with the empirical cumulative probability function obtained from sampling $1000$ Ginibre matrices of size $N=200$ and $N=1000$ respectively.}
  \label{fig:finiteNGinibre}
\end{figure}

Figure \ref{fig:finiteN} shows $F_N(Y)$ evaluated for increasing values of $N$. It is clear that for any practical applications, the limiting distribution is not a good approximation. However, the result presented in \eqref{eq:gammaratio} provides a closed expression for $F_N(Y)$ for finite $N$ in terms of the regularised incomplete gamma function which can easily be evaluated numerically by inserting the slowly varying function \eqref{eq:slowly} into \eqref{eq:cpf}, as had been done in Figure \ref{fig:finiteN}. For large values of $N$, one can benefit from the fact that the sum in \eqref{eq:cpf} is dominated by $n\approx N$. This is equivalent to the statement that the major contribution to the integral \eqref{eq:EM'} comes from the terms with $\xi$ close to one, as given by \eqref{eq:changev}. We can also verify the correctness of this expression by comparing the result of $F_N(Y)$ with the empirical cumulative probability function obtained from sampling $m$ representations of a Ginibre matrix of size $N$. This is shown in Figure \ref{fig:finiteNGinibre} for $m=1000$, with $N=200$ and $N=1000$ respectively.

\section{Finite $N$ corrections for the Gaussian potential} \label{sec:finite N}

One interesting aspect of the here presented approach is that it follows closely the standard finite $N$ expansion of the free energy of Hermitian random matrices using orthogonal polynomials (see for example Section 2.3 of \cite{Marino}). This provides a well-established framework for calculating finite $N$ corrections as we discuss in this section. Furthermore, besides this ``perturbative'' expansion, ``non-perturbative'' expansions, so-called instantons, can in principle be used to obtain large deviations (for a corresponding result for Unitary ensembles see \cite{Atkin}).

Calculating the next-to-leading order terms of $F_N(Y)$ can be important in practice since the convergence to the Gumbel distribution is very slow as can be seen from the order of the subleading term in \eqref{eq:largeN}, manifested in Figure  \ref{fig:finiteN}. We now compute the next-to-leading order terms in \eqref{eq:largeN} and \eqref{eq:finalGaussian} up to, but not including terms which are of order $1/\sqrt{N}$ and potential logarithmic factors (which is the first correction that is not slowly varying). Those are the most important corrections, since those logarithmic terms only fall off very slowly. As we will see later, there is an infinite number of such slowly varying correction terms.

Before presenting an explicit expression for all logarithmic correction terms, we outline the general procedure to obtain any higher-order correction. Firstly, we can extend \eqref{eq:EM} to any order using the Euler-MacLaurin formula
\bea
\mathrm{log}\big(F_N(y)\big)\!&=&\!N\!\int_0^1g(\xi;y)\,\mathrm{d}\xi+\frac{1}{2}\big[g(1;y)-g(0;y)\big]\!+ \nonumber \\
 && \quad + \sum_{p=1}^\infty \frac{1}{N^{2p}} \frac{B_{2p}}{(2p)!}\big[g^{(2p-1)}(1;y)-g^{(2p-1)}(0;y)\big], \label{eq:EMgeneral}
\eea
where $B_p$ are the Bernoulli numbers and
\beq
g(\xi;y) = \log\frac{h_{[\xi N]}(y)}{h_{[\xi N]}(\infty)}, \quad  h_n(y)=2\pi\,\int_0^y e^{Nf(r)}\,\mathrm{d}r
\eeq
Using that $f(r)=\frac{2\xi N+1}{N}\;\log r - V(r)$ (for general potential $V(r)$) and demanding that $f'(r_0) = 0$ which determines $r_0(\xi)$, we can expand
\beq \label{eq:allf}
f(r)=f(r_0) -\frac{1}{2}\big|f''(r_0)\big|\,(r-r_0)^2 + \sum_{p=3}^\infty \frac{1}{p!} f^{(p)} (r_0) (r-r_0)^p.
\eeq

From this we can perform a Feynman type expansion
\bea
h_n(y)&=& 2\pi\, e^{Nf(r_0)} \int_0^y e^{-\frac{N}{2}\big|f''(r_0)\big|\,(r-r_0)^2} \left[ 1 - N \sum_{p=3}^\infty \frac{1}{p!} f^{(p)} (r_0) (r-r_0)^p + \right. \nonumber \\
&&\quad \left. + \frac{1}{2}\left(- N \sum_{p=3}^\infty \frac{1}{p!} f^{(p)} (r_0) (r-r_0)^p\right)^2 + ... \right].
\eea
Each integral is Gaussian (given that $f''(r_0)<0$) and results in a term involving an error function which has an expansion
\beq\label{eq:erf2}
\erf(z)\equiv\frac{2}{\sqrt{\pi}}\,\int_0^z e^{-x^2}\,\mathrm{d}x=1-\frac{e^{-z^2}}{\sqrt{\pi}z}+\frac{e^{-z^2}}{2\sqrt{\pi}z^3}+....
\eeq
Inserting back each expansion in the previous one and carefully collecting terms of equal order one can obtain the general large $N$ expansion of $\log F_N(y)$.

Here we are only interested in logarithmic corrections for the Gaussian potential. We have already seen in the previous section that the second term in \eqref{eq:EMgeneral} is of order $\mathcal{O}(1/\sqrt{N \log N})$, given our choice of $y$, i.e.~\eqref{eq:yscalingrelation}. This is also true for terms coming from the Taylor expansion of $N f(r)$, i.e.~\eqref{eq:allf}, beyond the quadratic term. Thus we have that
\footnote{For the remainder of this section $+...$ refers to correction terms of the order of $\mathcal{O}(N^{-1/2})$ times some logarithmic function. In principle, we would be happy in ignoring anything which is going to zero faster than any slowly varying term, i.e. terms of $\mathcal{O}(N^{-\epsilon})$, for any $\epsilon>0$. It just happens that the most dominant of such terms is $\mathcal{O}(N^{-1/2})$ times some logarithmic function.}
\beq
\mathrm{log}\big(F_N(y)\big) = N \int_0^1g(\xi;y)\,\mathrm{d}\xi + ...,
\eeq
with $g(\xi;y)$ given in \eqref{eq:g}. For $y$ given in \eqref{eq:yscalingrelation}, $g(\xi;y)$ is dominated by the second error function in the numerator of \eqref{eq:g} which can be used to obtain, for the Gaussian potential,
\beq 
\mathrm{log}\big(F_N(y)\big)  = -\frac{N}{2}\int_0^1 \left[\erf\Big(\sqrt{2N}\,\big(y-r_0(\xi)\big)\Big) -1 \right]\,\mathrm{d} \xi + ...,
\eeq
where we expanded the logarithm to first order using the fact that higher order terms from the expansion of the logarithm in \eqref{eq:g} beyond the linear term go to zero at least as fast as $\mathcal{O}(1/\sqrt{N})$. This can furthermore, be written as:
\beq \label{eq:GaussianFiniteN1}
\mathrm{log}\big(F_N(y)\big)  = -\frac{N}{2}\int_0^1\erfc\Big(\sqrt{2N}\,\big(y-r_0(\xi)\big)\Big) \mathrm{d}\,\xi + ...,
\eeq
Using a change of variable $z\coloneqq\sqrt{2N}\,\big(y-r_0(\xi)\big)$ and the explicit form of $r_0(\xi) = \sqrt{\xi} +\mathcal{O}(1/N)$ for the Gaussian potential we can write
\beq 
\mathrm{log}\big(F_N(y)\big)  =\frac{1}{2}\int_{\sqrt{2N}\,(y-1)}^{\sqrt{2N}\,y}\,(z-\sqrt{2N}\,y)\,\mathrm{erfc}(z) \mathrm{d}z +... .
\eeq
\begin{figure}
  \centering
  \includegraphics[width=0.90\linewidth]{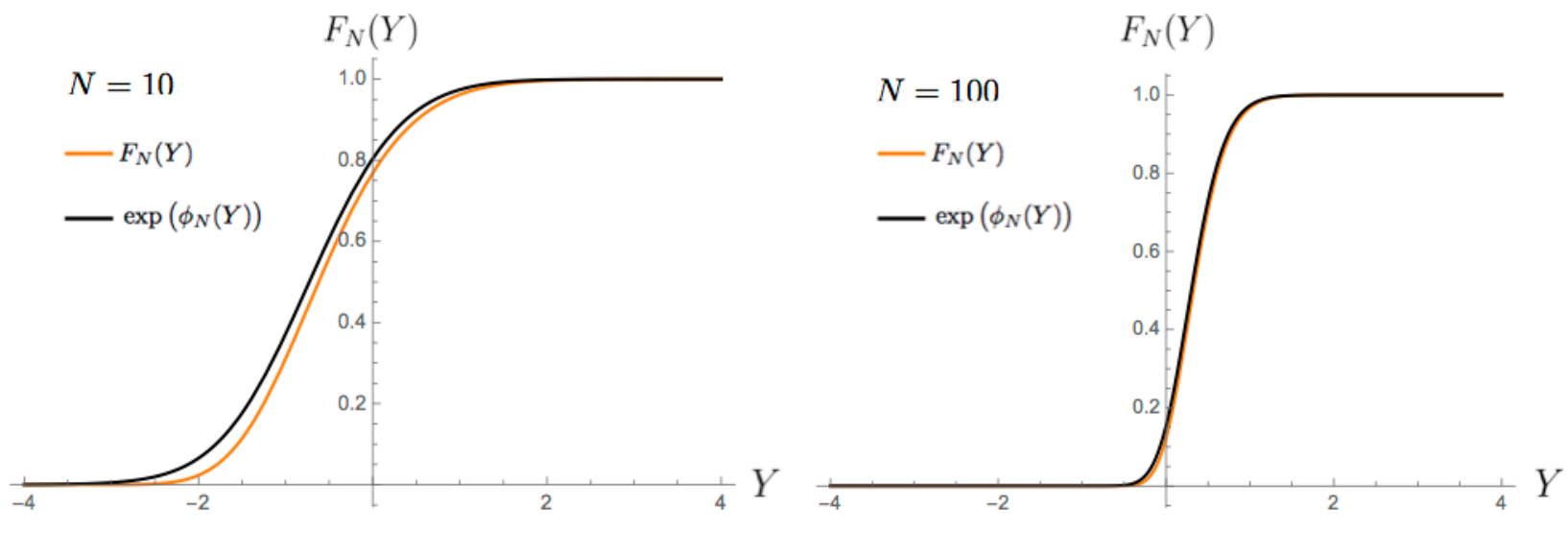}
  \caption{Shown is the function $F_N(Y)$ as well as the approximation $\exp(\phi_N(Y))$ for $N=10$ and $N=100$. One observe an excellent agreement for $N=100$.}
  \label{fig:finiteN2}
\end{figure}
In fact, one can integrate this expression which yields\footnote{Note that the leading order term cancels out.}
\bea
\log F_N(Y) =-\frac{1}{2}\left\{\left[\sqrt{2N}+\frac{\sigma_N(Y)}{2}\right]\frac{e^{-\sigma_N(Y)^2}}{\sqrt{\pi}} - \right.\nonumber \\
\quad \quad \quad \quad \, \, \left. -\erfc\left(\sigma_N(Y)\right)\left[\frac{\sigma_N(Y)^2}{2}+\sqrt{2N}\, \sigma_N(Y)+\frac{1}{4}\right]\right\} + \mathcal{O}(N^{-1/2}) \label{eq:finitefirstexpression}
\eea
where we substituted in the scaling relation \eqref{eq:yscalingrelation} for $y$ and defined
\beq
\sigma_N(Y) := \alpha_N+\frac{Y}{2\alpha_N} . 
\eeq
with $\alpha_N$ is given in \eqref{eq:slowly}. We observe that \eqref{eq:finitefirstexpression} still contains terms of order $\mathcal{O}(N^{-1/2})$, dropping those we get
\beq
\log F_N(Y) =  \phi_N(Y)+ ...
\eeq
with
\beq \label{eq:finiteNresult}
\phi_N(Y) := -\frac{\sqrt{2N}}{2}\left\{\frac{e^{-\sigma_N(Y)^2}}{\sqrt{\pi}} - \sigma_N(Y) \erfc\left(\sigma_N(Y)\right) \right\}.
\eeq
This provides an analytic expression including the leading order term of $\log F_N$, the Gumbel distribution, and all slowly varying correction terms! Figure \ref{fig:finiteN2} shows a plot of $F_N(Y)$ versus $\exp(\phi_N(Y))$ for $N=10$ and $N=100$. We see that already for $N=100$ the analytical expression $\exp(\phi_N(Y))$ matches the numerically evaluated $F_N(Y)$ perfectly. 

To see that \eqref{eq:finiteNresult} is indeed an infinite series of slowly varying terms in $N$, we use the asymptotic expansion of the complementary error function
\beq
\erfc(z) = \frac{e^{-z^2}}{\sqrt{\pi} z} \sum_{k=0}^\infty \frac{(-1)^k}{2^k}\frac{(2k-1)!!}{z^{2k}}, \quad \Re(z)>0
\eeq
and the definition of $\alpha_N$, given in \eqref{eq:slowly}, to write
\bea
\phi_N(Y) &=& - e^{-Y} \log (N ) e^{-\frac{Y^2}{4\alpha_N^2} }  \sum_{k=1}^\infty \frac{(-1)^{k+1}}{2^k}\frac{(2k-1)!!}{\sigma_N^{2k}(Y)} \nonumber \\
  &=& - e^{-Y} \left\{ 1  + \left( 2\log\log N  + \log 2\pi -3- 2Y- \frac{Y^2}{2} \right)\frac{1} {\log N} + \right. \nonumber \\
   && \quad\quad\quad\quad \,\,  \left. + \, \mathcal{O}\left( \frac{\log\log N}{\log^2 N}  \right)  \right\}. \label{eq:finiteNresultExpansion}
\eea
This highlights that \eqref{eq:finiteNresult} indeed constitutes the leading order term plus an infinite number of slowly varying correction terms. 

\section{Universality at the outer edge} \label{sec:outer}

In this section we consider the generalisation of the Gaussian potential to an arbitrary radially symmetric potential, $V\equiv V(r)$. The approach presented in Section \ref{sec:Gaussian} translates directly to an arbitrary potential by simply replacing $f(r)$ in \eqref{eq:hwithf} with
\beq
f(r)=\frac{2n+1}{N}\;\log r - V(r)
\eeq
The local extremum $r_0=r_0(\xi)$ determined by $f'(r_0) =0$ is given by the solution to the following equation
\beq\label{eq:r0}
r_0(\xi)\,V'\big(r_0(\xi)\big)=\frac{2n+1}{N} =2\xi + N^{-1}.
\eeq
Comparing \eqref{eq:r0} and \eqref{eq:outer} shows that in the large $N$ limit, the outer edge and $r_0$ coincide, i.e., $r_0(1)=a_++\mathcal{O}(N^{-1})$.

Next we Taylor expand $f(r)$ around its extremum at $r_0$. To ensure that the extremum is indeed a maximum we compute the second derivative,
\begin{eqnarray} \label{eq:fpp}
f''(r_0) & = & - \left(\frac{V'(r_0)}{r_0}+V''(r_0)\right) = -\left(\frac{1}{r}\,\frac{\mathrm{d}}{\mathrm{d}r}\Big(r\,V'(r)\Big)\right)\Bigg|_{r=r_0}.
\end{eqnarray}
Therefore, we have $f''(r_0)<0$ and thus have a maximum, if the condition
\beq\label{eq:condition1}
r\,V'(r)\mbox{  increasing in}\; \mathbb{R}^{+}
\eeq
or equivalently
\beq\label{eq:condition2}
V'(r)>0\;\mbox{and}\; V\; \mbox{convex in}\; \mathbb{R}^{+}
\eeq
holds for the potential $V(r)$ given the condition \eqref{eq:bounded}. Note that these conditions are the same as the ones given below \eqref{eq:bounded}.

Since we have shown that $f''(r_0)<0$ under the condition \eqref{eq:condition1} or \eqref{eq:condition2}, it then follows that \eqref{eq:hfromerf} and \eqref{eq:gintermediate} from the Gaussian case remain valid, where now $f''(r_0)$ in $b =| f''(r_0)/2|$ is given by \eqref{eq:fpp}. Furthermore, using a Taylor expansion around $\xi=1$ we get to leading order,
\beq \label{eq:generalb}
 b = \frac{ f''(r_0(1)) }{2} + ...=\frac{ F_+ }{2} + ... \quad \mbox{with}\quad F_+:= \frac{V'(a_+)}{a_+}+V''(a_+).
\eeq
Therefore, the leading order term in the probability distribution of the eigenvalue with the largest modulus is given by
\beq\label{eq:asymp}
N\,\int_0^1 g(\xi;y)\,\mathrm{d}\xi\sim-\frac{N}{2\sqrt{\pi}}\int_0^1 \frac{e^{-Nb\,(y-r_0(\xi))^2}}{\sqrt{Nb}\,\big(y-r_0(\xi)\big)}\,\mathrm{d}\xi,\hspace{.5cm}N\to\infty.
\eeq
As in the Gaussian case, we use the general scaling ansatz for $y$ given by \eqref{eq:yscalingrelation}, i.e. 
\beq
y= a_+ + \sqrt{\frac{2}{N F_+}}  \left( \alpha_N^{(+)} + \frac{1}{2\alpha_N^{(+)}}Y \right),
\eeq
where we replaced $b$ by \eqref{eq:generalb} and the slowly varying function $ \alpha_N^{(+)}$ will be determined later.

We thus also change the integration variable $\xi$ in \eqref{eq:asymp} as in \eqref{eq:changev} only with an additional constant $\gamma_+>0$, 
\beq
\xi=1-\gamma_+\,\frac{(NF_+/2)^{-1/2}}{2\alpha_N^{(+)}}\,X.
\eeq
The value of $\gamma_+$ will be determined in the following. We expand $r_0(\xi)$ around $\xi=1$:
\begin{eqnarray}
r_0(\xi) & = & r_0(1)+\mathrm{d}r_0(\xi)/\mathrm{d}\xi\big|_{\xi=1}\,(\xi-1)+\mathcal{O}\left((\xi-1)^2\right)\nonumber\\
& = & a_++\delta_+\,(\xi-1)+\mathcal{O}\left((\xi-1)^2\right),
\end{eqnarray}
where $\delta_+$ is given by 
\begin{eqnarray}
\delta_+ := \mathrm{d}r_0(\xi)/\mathrm{d}\xi\big|_{\xi=1} & = & \frac{2}{V'(a_+)+a_+V''(a_+)}\nonumber\\
& = & 2\left(\frac{\mathrm{d}}{\mathrm{d}r}\Big(rV'(r)\Big)\Big|_{r=a_+}\right)^{-1}.
\end{eqnarray}
Hence, according to the constraint already imposed in \eqref{eq:condition1}, we have $\delta_+>0$ and therefore obtain
\beq
\sqrt{Nb}\,\left(y-r_0(\xi)\right)=\alpha_N^{(+)} +\frac{1}{2\/\alpha_N^{(+)}}\,(\gamma_+\delta_+\,X+Y)+\mathcal{O}\left(N^{-1/2}/\alpha_N^{(+)}\right).
\eeq
which by choosing
\beq
\gamma_+=\frac{1}{\delta_+}=\frac{1}{2}\,\frac{\mathrm{d}}{\mathrm{d}r}\Big(r\,V'(r)\Big)\Big|_{r=a_+} = \frac{a_+ F_+}{2}
\eeq
reduces to \eqref{eq:errorarg}. Thus we have,
\begin{eqnarray}
N\,\int_0^1 g(\xi;y)\,\mathrm{d}\xi & = & -\frac{a_+\sqrt{NF_+}\,e^{-(\alpha_N^{(+)})^2}}{4\sqrt{2\pi}\,(\alpha_N^{(+)})^2}\,e^{-Y} +... .
\end{eqnarray}
The pre-factor is one to leading order if we choose,
\beq\label{eq:slowly2}
\alpha_N^{(+)}=\frac{1}{\sqrt{2}}\,\left|\log N-2\log\log N-\log2\pi  + \log\frac{a_+^2 F_+}{4} \right|^\frac{1}{2}.
\eeq 
For the Gaussian potential the last term in the above expression is zero.

We have thus shown universality of the distribution of the eigenvalue with the largest modulus when rescaled around the outer edge of a generic potential $V(r)$ satisfying the condition given in \eqref{eq:condition1} or \eqref{eq:condition2}, where the limiting probability distribution is the Gumbel distribution
\beq
\lim_{N\rightarrow\infty}\mathbb{P}_N\left(|z_{max}|\le a_+ +  \sqrt{\frac{2}{N F_+}} \left[\alpha_N^{(+)}+\frac{Y}{2\alpha_N^{(+)}} \right]\right)=e^{\,-e^{-Y}},
\eeq
with $\alpha_N^{(+)}$ given in \eqref{eq:slowly2} and $F_+$ in \eqref{eq:generalb}.
This reproduces results from \cite{Chafai} using our general approach. 

\section{Universality at the inner edge} \label{sec:inner}

Now we study the extreme value statistics of a normal matrix ensemble with potential $V=V(r)$ at the inner edge of the eigenvalue support for potentials where $a_->0$. 
To do so we now choose $\mathcal{D}$ in \eqref{eq:genZ} to be $\mathcal{D}=\{ z : \, |z|\geq y\}$ and define
\beq
Z_N(y):=Z_N(\{ z : \, |z|\geq y\}).
\eeq
The probability distribution for the eigenvalue with the smallest modulus $z_{min}$ is then given by
\beq\label{eq:probability2}
F_N(y) := \mathbb{P}_N(|z_{min}|\ge y)=\frac{Z_N(y)}{Z_N(0)}=\prod_{n=0}^{N-1}\frac{h_n(y)}{h_n(0)},
\eeq
where now $h_n(y)$ is given through
\beq
\int e^{-N\,V(z)}\,\mathbb{I}_{\{|z|\geq y\}}\,p_n(z;y)\,\overline{p_m(z;y)}\,\mathrm{d}^2z=h_n(y)\,\delta_{nm}.
\eeq
We want to evaluate the above expression and scale it around the finite inner edge of the support at $|z|=a_->0$. We have
\beq
h_n(y)=2\pi\,\int_y^{\infty} r^{2n+1}\,e^{-N V(r)}\,\mathrm{d}r=2\pi\,\int_y^{\infty} e^{Nf(r)}\,\mathrm{d}r
\eeq
with $f(r)=(2n+1)/N\;\log r-V(r)$. Then performing a Taylor expansion and following a similar strategy as in the previous section results in
\beq
\frac{h_n(y)}{h_n(0)}=\frac{1+\erf(\sqrt{Nb}\,(r_0-y))}{1+\erf(\sqrt{Nb}\,r_0)} + ...,
\eeq
where again $b\equiv\big|f''(r_0)/2\big|$ and $r_0$ is given as a solution of \eqref{eq:r0}. Thus in the large $N$ limit one has again \eqref{eq:EM}, i.e.,
\beq\label{eq:EM2}
\mathrm{log}\big(F_N(y)\big)=N\int_0^1 g(\xi;y)\,\mathrm{d}\xi+\frac{1}{2}[g(1;y)-g(0;y)]+\mathcal{O}\left(N^{-1}\right),
\eeq
but where now
\beq \label{eq:EM2b}
g(\xi;y) = \log\left(\frac{1+\erf\Big(\sqrt{Nb}\,\big(r_0(\xi)-y\big)\Big)}{1+\erf\Big(\sqrt{Nb}\,r_0(\xi)\Big)}\right) + ... .
\eeq
which using a reasoning similar to the one in the previous sections yields
\beq\label{eq:EMleading}
N\,\int_0^1 g(\xi;y)\,\mathrm{d}\xi\sim-\frac{N}{2\sqrt{\pi}}\int_0^1 \frac{e^{-Nb\,(r_0(\xi)-y)^2}}{\sqrt{Nb}\,\big(r_0(\xi)-y\big)}\,\mathrm{d}\xi,\hspace{.5cm}N\to\infty,
\eeq
where now to leading order we expand around $\xi=0$,
\beq\label{eq:generalb2}
 b =  \frac{ |f''(r_0(0))|}{2} + ... =\frac{ F_-}{2} + ...  \quad \mbox{with}\quad F_- :=  V''(a_-).
\eeq

In fact, the argument is now even simpler since $r_0(\xi)$ stays finite and does not go to zero as $\xi$ goes to zero -- recall that $V'(r_0(\xi))$ goes to zero when $\xi$ goes to zero and thus \eqref{eq:r0} does not require that $r_0(\xi)$ becomes zero. This is assured by the requirement that $a_->0$.

Now we scale $y$ around the inner edge of the support, $a_-$, in an analogous way as for the outer edge
\beq
y=a_-- \sqrt{\frac{2}{N F_-}}  \left(\alpha_N^{(-)} + \frac{Y}{2\alpha_N^{(-)}} \right).
\eeq
Note that as $N$ becomes large $y$ approaches $a_-$. Moreover, $r_0(\xi)$ with $\xi=0$ also approaches $a_-$ as $N$ becomes large, as is clear when comparing \eqref{eq:inner}
with \eqref{eq:r0}. This implies that the leading order contribution of \eqref{eq:EMleading} now comes from $\xi = 0$. This suggests the following change of variables in \eqref{eq:EMleading}
\beq
\xi=\gamma_-\,\frac{(NF_-/2)^{-1/2}}{2\alpha_N^{(-)}}\,X.
\eeq
Now expanding $r_0(\xi)$ around $\xi=0$ gives

\begin{eqnarray}
r_0(\xi) & = & r_0(0)+\mathrm{d}r_0(\xi)/\mathrm{d}\xi\big|_{\xi=0}\,\xi+\mathcal{O}(\xi^2)\nonumber\\
& = & a_-+\delta_-\,\xi+\mathcal{O}(\xi^2),
\end{eqnarray}
where recalling \eqref{eq:r0} and $V'(a_-)=0$ one finds that $\delta_-$ is given by
\beq
\delta_-\equiv\mathrm{d}r_0(\xi)/\mathrm{d}\xi\big|_{\xi=0}=\frac{2}{a_-V''(a_-)}.
\eeq
We notice that since $a_-$ is finite and the potential $V$ is convex, we have $\delta_->0$. We therefore obtain
\beq
\sqrt{Nb}\,\big(r_0(\xi)-y\big)=\alpha_N^{(-)}+\frac{1}{2\alpha_N^{(-)}}\,(\gamma_-\delta_-\,X+Y)+\mathcal{O}\left(N^{-1/2}/\alpha_N^{(-)}\right),
\eeq
which by choosing
\beq
\gamma_-=\frac{1}{\delta_-}=\frac{a_-\,V''(a_-)}{2} = \frac{a_- F_-}{2}
\eeq
becomes
\beq
\sqrt{Nb}\,\big(r_0(\xi)-y\big)=\alpha_N^{(-)}+\frac{1}{2\alpha_N^{(-)}}\,(X+Y)+\mathcal{O}\left(N^{-1/2}/\alpha_N^{(-)}\right).
\eeq
We thus have
\beq
N\,\int_0^1 g(\xi;y)\,\mathrm{d}\xi = -\frac{a_-\sqrt{NF_-}\,e^{-(\alpha_N^{(-)})^2}}{4\sqrt{2\pi}\,(\alpha_N^{(-)})^2}\,e^{-Y} +... .
\eeq
Similar to the previous section, the pre-factor is one to leading order, if we choose,
\beq\label{eq:slowly3}
\alpha_N^{(-)} =\frac{1}{\sqrt{2}}\,\left|\log N-2\log\log N-\log2\pi  + \log\frac{a_-^2 F_-}{4} \right|^\frac{1}{2}.
\eeq 

Following the reasoning from the previous section we thus get
\beq
F(Y)= \lim_{N\rightarrow\infty}\mathbb{P}_N\left(|z_{min}|\ge a_-   - \sqrt{\frac{2}{N F_-}}   \left[\alpha_N^{(-)}+\frac{Y}{2\alpha_N^{(-)}} \right] \right)=e^{\,-e^{-Y}},
\eeq
where $\alpha_N^{(-)}$ is given in \eqref{eq:slowly3} and $F_-$ in \eqref{eq:generalb2}. This shows universality by proving convergence of the rescaled distribution of the eigenvalue with the smallest modulus to a Gumbel distribution. The derivation assumes that the potential fulfills the condition \eqref{eq:condition1} or \eqref{eq:condition2} and that the inner radius $a_->0$, i.e., the support of the eigenvalue density has topology of a ring.

\section{Finite $N$ corrections for outer and inner edge of an arbitrary potential} \label{sec:finiteNgen}

To obtain all logarithmic corrections for the extreme value statistics at the outer edge of an arbitrary potential we start with the generalisation of  \eqref{eq:GaussianFiniteN1},
\beq
\mathrm{log}\big(F_N(y)\big)  = -\frac{N}{2} \int_0^1\erfc\Big(\sqrt{\frac{N|f''(r_0)|}{2}}\,\big(y-r_0(\xi)\big)\Big) \mathrm{d}\,\xi + ....
\eeq
We can now change the integration to $r_0$, where the change in the integration measure can be obtained from \eqref{eq:r0},
\beq
\mathrm{log}\big(F_N(y)\big)  = -\frac{N}{4} \int_{r_0(0)}^{r_0(1)} \erfc\Big( \sqrt{\frac{N|f''(r_0)|}{2} }\,\big(y-r_0\big)\Big) r_0 |f''(r_0)| \mathrm{d} r_0 + ....
\eeq
with 
\beq
|f''(r_0)| = \frac{V'(r_0)}{r_0}+V''(r_0)
\eeq
thus eliminating any $\xi$-dependence. Next we change the integration variable to 
\beq
z(r_0) = \sqrt{\frac{N|f''(r_0)|}{2}} (y-r_0)
\eeq
Knowing that for our scaling of $y$ the leading contribution from the error function comes from $r_0=a_+$ we Taylor expand the measure around this value, leading to
\beq
\mathrm{log}\big(F_N(y)\big)  = \frac{1}{2} \int_{z(a_-)}^{z(a_+)} \left(  z- y \sqrt{\frac{N|f''(a_+)|}{2}}   \right)  \erfc( z ) \mathrm{d} z + ....
\eeq
where in addition we substituted $r_0(1)=a_++...$ and $r_0(0)=a_-+...$. Performing the integration, inserting the scaling for $y$, and following the step of Section \ref{sec:finite N}, we arrive at
\beq
\log F_N(Y) =  \phi_N^{(+)}(Y)+ ...
\eeq
with
\beq \label{eq:finiteNresultOuter}
\phi_N^{(+)}(Y) := -\frac{a_+ \sqrt{N F_+ } }{2\sqrt{2}}\left\{\frac{e^{-\sigma_N^{(+)}(Y)^2}}{\sqrt{\pi}} - \sigma_N^{(+)}(Y) \erfc\left(\sigma_N^{(+)}(Y)\right) \right\}.
\eeq
Here we defined, as previously, 
\beq
\sigma_N^{(+)}(Y) := \alpha_N^{(+)}+\frac{Y}{2\alpha_N^{(+)}} ,
\eeq
but with $\alpha_N^{(+)}$ given by \eqref{eq:slowly2} and $F_+$ is defined in \eqref{eq:generalb}.  The above expression provides an infinite series of all logarithmic correction terms to the extreme value statistics of the largest eigenvalue scaled around the outer edge of the eigenvalue support of an arbitrary radially symmetric potential. We see that only the first two derivatives of the potential, $V'(r)$ and $V''(r)$, at the outer edge enter the expression, but not higher order derivatives. 

As for the Gaussian case, we can also give a series expression for \eqref{eq:finiteNresultOuter}, highlighting that it is an infinite series of slowly varying terms,
\bea
\phi_N^{(+)} (Y)&=& - e^{-Y} \log (N ) \exp\left(-\frac{Y^2}{4(\alpha_N^{(+)})^2} \right) \sum_{k=1}^\infty \frac{(-1)^{k+1}}{2^k}\frac{(2k-1)!!}{[\sigma_N^{(+)}(Y)]^{2k}} \nonumber \\
  &=& - e^{-Y} \left\{ 1  + \left( 2\log\log N  + \log \frac{8\pi}{a_+^2 F_+} -3- 2Y- \frac{Y^2}{2} \right)\frac{1} {\log N} + \right. \nonumber \\
   && \quad\quad\quad\quad \,\,  \left. + \, \mathcal{O}\left( \frac{\log\log N}{\log^2 N}  \right)  \right\}, \label{eq:infiniteexp}
\eea
where the only change is in the form of the slowly varying function.

\begin{figure}
  \centering
  \includegraphics[width=0.70\linewidth]{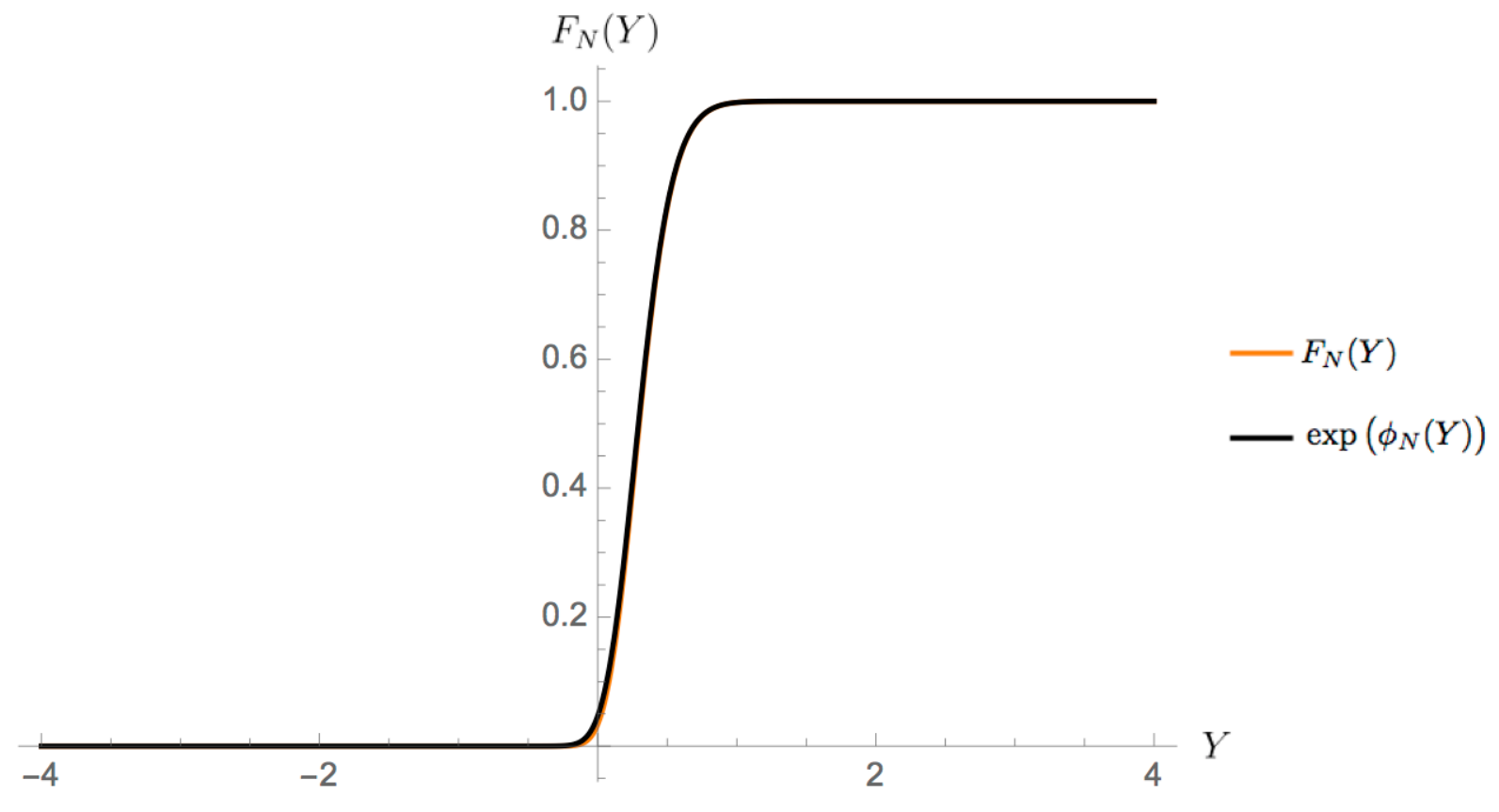}
  \caption{Shown is the function $F_N(Y)$ for the outer edge as well as the approximation $\exp(\phi_N^{(+)}(Y))$ for $N=100$ evaluated for the potential $V(r) = r^3/3$.}
  \label{fig:finiteNGen1}
\end{figure}

As an example we look at the potential $V(r) = r^3/3$. For this potential, $a_+ = 2^{1/3}$ and $F_+ = 3 a_+$. In Figure \ref{fig:finiteNGen1} we plot for this potential $\exp(\phi_N^{(+)}(Y))$, with $\phi_N^{(+)}(Y)$ given in \eqref{eq:finiteNresultOuter}, against the numerically evaluated $F_N(Y)$ for $N=100$.

The derivation for the inner edge proceeds analogously. Firstly, starting with \eqref{eq:EM2} and \eqref{eq:EM2b} and proceeding as in the previous case, we obtain for $F_N(y) := \mathbb{P}_N(|z_{min}|\ge y)$,
\beq
\mathrm{log}\big(F_N(y)\big)  = -\frac{N}{2} \int_0^1\erfc\Big(\sqrt{\frac{N|f''(r_0)|}{2}}\,\big(r_0(\xi) - y \big)\Big) \mathrm{d}\,\xi + ....
\eeq

The derivation now proceeds similarly as for the outer edge, where now, due to the different scaling of $y$, the main contribution of the error function comes from $\xi=0$ which corresponds to $r_0 = a_-$. Proceeding as above one obtains,
\beq
\log F_N(Y) =  \phi_N^{(-)}(Y)+ ...
\eeq
with
\beq \label{eq:finiteNresultInner}
\phi_N^{(-)}(Y) := -\frac{a_- \sqrt{N F_- } }{2\sqrt{2}}\left\{\frac{e^{-\sigma_N^{(-)}(Y)^2}}{\sqrt{\pi}} - \sigma_N^{(-)}(Y) \erfc\left(\sigma_N^{(-)}(Y)\right) \right\}.
\eeq
as well as 
\beq
\sigma_N^{(-)}(Y) := \alpha_N^{(-)}+\frac{Y}{2\alpha_N^{(-)}} ,
\eeq
with $\alpha_N^{(-)}$ for given by \eqref{eq:slowly3} and $F_-$ is defined in \eqref{eq:generalb2}.

\begin{figure}
  \centering
  \includegraphics[width=0.70\linewidth]{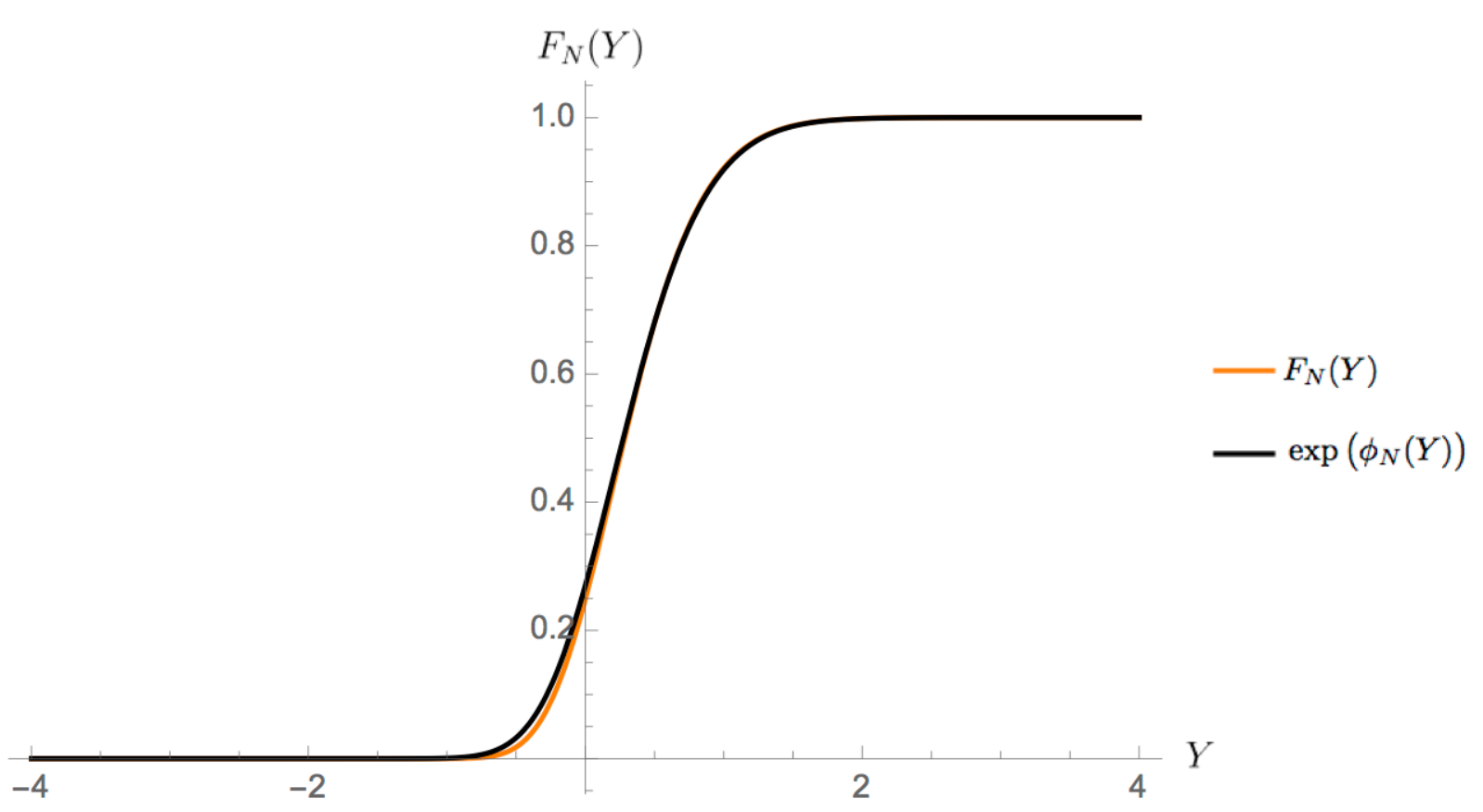}
  \caption{Shown is the function $F_N(Y)$ for the inner edge as well as the approximation $\exp(\phi_N^{(-)}(Y))$ for $N=500$ evaluated for the potential $V(r) = r^2/2 - r$.}
  \label{fig:finiteNGen2}
\end{figure}

The infinite expansion of $\phi_N^{(-)}(Y)$ in terms of logarithmic corrections is exactly as in \eqref{eq:infiniteexp} with all plus sub/super-scripts replaced by minus sub/super-scripts.

As an illustration for how well $\exp(-\phi_N^{(-)}(Y))$ approximates the finite $N$ expression $F_N(Y)$ we plot in Figure \ref{fig:finiteNGen2} both for the potential $V(r)=r^2/2 - r$ for $N=500$. For this potential the inner edge is located at $a_-=1$ and $F_-=1$.

\section{Discussion} \label{sec:Discussion}

The celebrated Tracy-Widom distribution provides the extreme value statistics for Hermitian random matrices. For non-Hermitian matrices the situation is somewhat easier as eigenvalues are only weakly correlated and the extreme value statistics is given by the much simpler Gumbel distribution. However, due to the lack of symmetries, non-Hermitian random matrices are in general harder to deal with. A special case are normal random matrices which are non-Hermitian and at the same time allow for a Coulomb gas formulation for general potential. 

In this work we investigate the extreme value statistics of normal random matrices and 2D Coulomb gases for general radially symmetric potentials. This is done by extending the orthogonal polynomial approach, introduced in \cite{Nadal} for Hermitian matrices, to normal random matrices and 2D Coulomb gases. We first analyse the simplest case of Gaussian normal random matrices and show convergence of the eigenvalue with the largest modulus rescaled around the outer edge of the eigenvalue support to a Gumbel distribution. One strength of this approach lies in the fact that it immediately generalises to an arbitrary potential $V=V(r)$ with radial symmetry which satisfies the condition \eqref{eq:bounded} together with the condition \eqref{eq:condition1} or  \eqref{eq:condition2}. We use this to show universality of the distribution of the eigenvalue with largest modulus when rescaled around the outer edge of the eigenvalue support. This provides an alternative, simplified derivation of results presented in \cite{Rider,Chafai}. In addition, it is shown that the approach presented here also generalised to compute convergence of the distribution of the eigenvalue with smallest modulus rescaled around the inner edge of the eigenvalue support with topology of an annulus. 

As noted in the introduction, it is interesting that one can extend the orthogonal polynomial approach developed for Hermitian random matrices to non-Hermitian random matrices. Furthermore, it is surprising that one can use the same derivational steps to obtain the extreme value statistics of normal random matrices as are used to obtain the Tracy-Widom distribution for Hermitian random matrices \cite{Nadal, Akemann}. This universal underlying structure suggests that the here presented approach can easily be generalised to new cases, as we indeed show when calculating the extreme values statistics for the eigenvalue with the smallest modulus for the ring distribution.

The here presented approach can also be used to obtain finite $N$ corrections. In  \eqref{eq:finiteNresult} we give an analytical expression for the leading order term and all slowly varying finite $N$ corrections for the Gaussian potential. It is quite interesting that one can obtain such a closed-form expression given that there is an infinite number of such slowly varying correction terms as one can see from \eqref{eq:finiteNresultExpansion}. Such corrections are important in practice given the extremely slow convergence of the extreme value statistics $F_N(Y)$, as illustrated in Figure \ref{fig:finiteN} which can be contrasted with the extremely good approximation given by \eqref{eq:finiteNresult} and illustrated in Figure \ref{fig:finiteN2} for much smaller $N$. In \eqref{eq:finiteNresultOuter} and \eqref{eq:finiteNresultInner} we obtain analogous results for the outer and inner edge of an arbitrary radially symmetric potential. The logarithmic corrections only depend on the first and second moment of the potential evaluated at the corresponding edge, but not on higher moments. Figure \ref{fig:finiteNGen1} and \ref{fig:finiteNGen2} show how well the expression including all logarithmic correction terms approximates the finite $N$ expression for two examples of potentials both at the outer as well as at the inner edge.

An interesting direction of future research is to explore the extreme value statistics of the eigenvalue with smallest modulus at the transition when the inner radius of the eigenvalue support goes to zero as $N$ goes to infinity. This transition was analysed in \cite{Cunden} in the context of the mean radial displacement. It would be interesting to see whether the extreme value statistics can be tuned to a different universality class in this transition by using a double scaling limit in which $a_-$ goes to zero as $N$ becomes large. \\

{\emph{Acknowledgements --}} The authors would like to thank John Wheater, Max Atkin and Carlos Tomei for fruitful discussions, as well as the anonymous referee for constructive comments. RE is thankful for the financial support of CCPG (PUC-Rio). SZ was previously supported by Nokia Technologies and Lockheed Martin through the Quantum Optimisation and Machine Learning Project at Oxford University.

\appendix

\section{Numerical methods} \label{sec:appA}

\subsection{Ginibre matrices}

For the Gaussian case we can generate samples from the joint eigenvalues density by first creating a complex matrix of random elements
\beq
M_{ij} = \frac{1}{\sqrt{2N}}\left(A_{ij} + i B_{ij}\right),\quad \mbox{with}\quad A_{ij}, B_{ij} \sim \mathcal{N}(0,1)
\eeq
and then use standard techniques to numerically diagonalise the matrix.

As a concrete example in {\tt R} one can write the following code, which generates $m$ samples of the matrix
$M_{ij}$, diagonalises it, extracts the eigenvalue with the largest modulus and stores it in an array $\vec{y}$.

\begin{verbatim}
y <- rep(NULL, m)
for (i in 1:m){
  M = matrix(complex(real=rnorm(N*N),imaginary=rnorm(N*N)),ncol=N)
  y[i] = sort(Mod(eigen(M, symmetric=FALSE,
           only.values = TRUE)$values)/sqrt(N*2))[N] }
\end{verbatim}

\subsection{Monte-Carlo methods for normal random matrices and Coulomb gases}

To obtain samples of the eigenvalues $\vec{z}$ from the joint eigenvalue density \eqref{eq:Pmeasure} for a general (radially symmetric) potential
we employ Monte-Carlo techniques. To do so we decompose $\vec{z} =\vec{x}+i\vec{y}$ and initialise the $N$ elements of $\vec{x}$ and $\vec{y}$ with
uniform random numbers in the interval $(-1,1)$. Then we iterate over the following steps. First one perturbs a given randomly chosen element of $\vec{x}$ and $\vec{y}$ each by a Gaussian random
number, multiplied by a scaling $\eta/N$ where $\eta$ is a number of $\mathcal{O}(1)$. Given the perturbed vector $\pvec{z}'$ we then evaluate  the difference in
energy in the Boltzmann weight \eqref{eq:Boltzmann},
\beq
\Delta E = N^2 (S_{\mathrm{eff}}(\vec{z}) - S_{\mathrm{eff}}(\pvec{z}')).
\eeq
If the new configuration has lower energy, i.e., $\Delta E >0$, we change $\vec{z}$ to $\pvec{z}'$. If the new configuration has larger energy we still allow to
change $\vec{z}$ to $\pvec{z}'$ if $\exp(\Delta E)$ is larger than a random uniform number $u$ in $(0,1)$. This is the standard Metropolis step.
The above iteration is repeated $m$ times where $m$ is usually chosen $\mathcal{O}(N^3)$.

For completeness we provide the pseudo-code of the algorithm below.

\begin{algorithmic}[1]
\Procedure{Monte Carlo simulation for eigenvalue distribution} {}
\State initialise: $N, m, \eta$ and the function $V(r)$
\State initialise: $\vec{x} \leftarrow  \mathrm{uniform}(-1,1)$
\State initialise: $\vec{y} \leftarrow  \mathrm{uniform}(-1,1)$
\For{interation in $1,.., m$}
	\State // perturb one randomly chosen element of $\vec{x}$ and $\vec{y}$:
	\State $k\leftarrow \mathrm{choose}([1,...,N])$ 
	\State $x'[k] \leftarrow  x[k] + \eta \delta_x/N$ with $\delta_x\sim \mathrm{normal}(0,1)$
	\State $y'[k] \leftarrow  y[k] + \eta \delta_y/N$ with $\delta_y\sim \mathrm{normal}(0,1)$
	\State // Calculate energy difference for $\vec{z}= \vec{x}+i \vec{y}$ and $\pvec{z}'= \pvec{x}'+i \pvec{y}'$:
	\State $\Delta E = N^2 (S_{\mathrm{eff}}(\vec{z}) - S_{\mathrm{eff}}(\pvec{z}'))$
	\State // sample a uniform number and do the Metropolis step:
	\State $u\sim \mathrm{uniform}(0,1)$
 	\If{$\exp(\Delta E) > u$}
 		\State $\vec{x} \leftarrow \pvec{x}'$ 
		\State $\vec{y} \leftarrow \pvec{y}'$
	\EndIf
\EndFor
\EndProcedure
\end{algorithmic}

\end{document}